\documentclass[a4paper,fleqn,usenatbib]{mnras}

\usepackage{amsmath}

\usepackage[T1]{fontenc}
\usepackage{ae,aecompl}

\usepackage{graphicx}	
\usepackage{amsmath}	
\usepackage{amssymb}	

\title[Bubbles, potential flows and stellar convection.]{On the
  relevance of bubbles and potential flows for stellar convection}

\author[M. M. Miller Bertolami et al.]{
M. M. Miller Bertolami,$^{1,2}$\thanks{E-mail: marcelo@mpa-garching.mpg.de}
M. Viallet,$^{1}$
V. Prat$^{1}$,
W. Barsukow$^{3}$
and A. Weiss$^{1}$
\\
$^{1}$Max-Planck-Institut f\"ur Astrophysik, Karl-Schwarzschild-Str. 1, 85748, Garching, Germany.\\
$^{2}$Instituto de Astrof\'isica de La Plata, UNLP-CONICET,
    Paseo del Bosque s/n, 1900 La Plata, Argentina.\\
$^{3}$Institut f\"ur Mathematik, Universit\"at W\"urzburg, Campus Hubland Nord,
Emil-Fischer-Strasse 40, 97074 W\"urzburg, Germany.}

\date{Accepted 2016 January 20. Received 2016 January 20; in original form 2015 November 06}

\pubyear{2015}

\begin{document}
\label{firstpage}
\pagerange{\pageref{firstpage}--\pageref{lastpage}}
\maketitle

\begin{abstract}
  Recently Pasetto et al.  have proposed a new method to derive a
  convection theory appropriate for the implementation in stellar evolution
  codes. Their approach is based on the simple physical picture of spherical
  bubbles moving within a potential flow in dynamically unstable regions, and a
  detailed computation of the bubble dynamics.
  Based on this approach the authors derive a new theory of convection which
  is claimed to be parameter free, non-local and time-dependent. This is a
  very strong claim, as such a theory is the holy grail of stellar
  physics. Unfortunately we have identified several distinct problems in the
   derivation which ultimately render their theory inapplicable to any
  physical regime. In addition we show that the framework of spherical bubbles
  in potential flows is unable to capture the essence of stellar convection,
  even when equations are derived correctly.

\end{abstract}




\section{Introduction}
It is not an exaggeration to state that the turbulent transport of heat,
angular momentum and chemical species is the most important unsolved problem
in stellar astrophysics. Most of the present uncertainties in stellar physics
are, in one way or another, linked to our incomplete understanding of mixing
in stellar interiors, e.g. the final fate of stars of high and intermediate
mass, formation of s-process elements, chemical anomalies on the red giant
branch, formation of carbon stars, size of the convective cores in H- and
He-burning stars. In spite of many decades of attempts to derive an accurate
time-dependent and non-local theory of convection that can be included in
stellar evolution codes, success has been very minor. While some theories of
time-dependent convection have been derived and applied
\citep{1986A&A...160..116K,kuhfuss-thesis,1998A&A...340..419W,flaskamp-thesis},
they all introduce several free parameters that must be calibrated for
different regimes, diminishing their predictive power. Even more problematic
is the case of non-local convection and convective boundary mixing. For
decades, serious attempts have been made to derive non-local convection
theories that could be introduced in stellar evolution codes
\citep{2006ApJ...643..426D,2008IAUS..252...83D,2011A&A...528A..80C}. However,
these theories are not popular due to their complexity and their limited
accuracy \citep{1986A&A...167..239X, 2007IAUS..239..314W}. Attempts to derive
a general framework for the treatment of stellar convection and other mixing
processes lead to very complex equations which cannot be easily included in 1D
stellar evolution codes
\citep{1986A&A...160..116K,2011A&A...528A..76C,2011A&A...528A..77C,2011A&A...528A..78C,2011A&A...528A..79C,2011A&A...528A..80C}. In
fact, non-local consequences of convection, such as convective boundary
mixing, are routinely included in stellar evolution codes based on ad-hoc
prescriptions and additional free parameters ---see \cite{2015A&A...580A..61V}
and \cite{2015ApJ...809...30A} for recent discussions on these
issues. Consequently, despite its well-known shortcomings, the mixing-length
theory (MLT; \citealt{Prandtl1925,1932ZA......5..117B,1953ZA.....32..135V})
has been in use for more than 80 years.

In this paper, we call a ``theory of stellar convection" a theory that can be
implemented in 1D stellar evolution codes. Such a theory should capture the
essential properties of turbulent convection, allowing to reproduce its
effects on the stellar structure (mainly chemical mixing and energy transport)
without the need to resort to expensive 3D simulations. A theory of stellar
convection is highly sought, as the predictive power of current stellar models
is strongly limited by the shortcomings of MLT. The status on the theoretical
side contrasts with the progress done in observational techniques and
instrumentation (e.g. KEPLER, CoRoT and GAIA; \citealt{2012Ap&SS.341...31D,
  2014IAUS..301.....G}). A new generation of stellar models is necessary to
fully exploit the large amount of quality data that is delivered by
observers. Without any doubt a new generation of stellar models should rely on
a better treatment of convection.

Recently, \cite{2014MNRAS.445.3592P} claimed to present an accurate
parameter-free, non-local, time-dependent theory of stellar convection that
can be easily implemented in 1D stellar codes.  This is a strong claim since
such a theory has been sought for many decades.  In order to facilitate the
reader's understanding, we start by schematically summarizing the method
proposed by \cite{2014MNRAS.445.3592P}. \cite{2014MNRAS.445.3592P} adopt a
rather simple picture of convection, in which the transport of heat is
achieved by ``bubbles'' that rise due to buoyancy in a convectively unstable
region. This description of convection using the concept of bubbles is likely
inspired from the usual simple picture that one has in mind when deriving the
MLT. Furthermore, in the picture of \cite{2014MNRAS.445.3592P}, convective
bubbles have a definite shape (they are spherical) and are differentiated from
the surrounding material ---i.e. the surrounding material flows around them.
As a first step the authors analyze the motion of an isolated bubble. From
kinematic considerations they derive the expression of the velocity field
around the bubble. This is done assuming that the flow around the bubble is a
potential flow ($\mathbf{\nabla} \times \mathbf{u} = 0$, where $\mathbf{u}$ is
the velocity field).  This allows them to link the velocity of the moving
bubble to that of the surrounding fluid at each time ---i.e. assuming an
instantaneous adjustment of the surrounding fluid. Given the velocity field
around the bubble, the authors then deduce the pressure field around the
bubble. Knowing the pressure field, they compute the total force that the
fluid exerts on the bubble,
$$
\vec F = \oint_\mathrm{bubble} P \vec n \mathrm{d}S.\nonumber
$$
Applying Newton's law to the bubble, the authors derive an expression for the
acceleration of the bubble, the first key result of their theory. With
appropriate initial conditions, this equation defines completely the motion of
the bubble as a function of time. In the second part of their work,
\cite{2014MNRAS.445.3592P} use their theory for an isolated bubble to
formulate a theory of convection by considering a collective set of
bubbles.

In this work we study the applicability of this method to the stellar regime
and its possible limitations. In order to do this we derive the equation for
the dynamics of the bubble by a careful accounting of the physical assumptions
and hypothesis made in the derivation. During this process we found that some
inconsistent physical and mathematical assumptions have been made by
\cite{2014MNRAS.445.3592P}, casting serious doubts on the validity of their
theory. It will also become clear in the next section that the claim of a
non-local, time-dependent theory is an overstatement by the authors. Yet, the
method of deriving a parameter-free convection theory from the full dynamics
of a convective element assuming a surrounding potential flow is
interesting. If valid, the method could indeed be extended to obtain a
parameter-free, non-local and time-dependent theory and to get rid of the
mixing length parameter $\alpha$ whose calibration in different stellar
regimes is problematic
\citep{1999A&A...346..111L,2014MNRAS.445.4366T,2015ApJ...799..142T,2015A&A...573A..89M}.
We have been able to reobtain the dynamics of the bubble by a sound
mathematical and physical derivation. This allow us to study the behavior of
the solutions and assess the physical regime in which the method described by
\cite{2014MNRAS.445.3592P} can be applied. Unfortunately we find that the
movement of spherical bubbles within potential flows is completely
inapplicable to the regime of stellar convection and that no useful theory of
stellar convection can be obtained from this approach. This is not a surprise
since the adoption of the ideal fluid and the potential flow approximations
(the ``dry water'' approximation, \citealt{1964flp..book.....F}) neglect the
importance of viscosity and boundary layers for the dynamics of the bubble.

The paper is organized as follows. In the next section we show that the
derivation of \cite{2014MNRAS.445.3592P} of the acceleration equation is
flawed due to incorrect physical and mathematical assumptions. In section
\ref{eq_motion} we clarify the approximations underlying their theory of an
isolated bubble, and provide the correct derivation of the acceleration
equation of the bubble.  We show that the authors misinterpreted their
acceleration equation, and neglected a term that is physically important. In
section \ref{sol_motion} we provide the correct analysis of the equation
of motion, and focus particularly on the asymptotic/final regimes reached by
the bubble. We show that it is unavoidable that the theory becomes
inconsistent and highly non-physical. In section \ref{conclusion}, we finish the
article with some discussion and concluding remarks.

\section{Physical and mathematical inconsistencies in
  Pasetto et al. (2014)}
\label{mistakes}

Before analyzing the physical and mathematical assumptions adopted by
\cite{2014MNRAS.445.3592P}, it is already worth noting that a first
consequence of the method adopted by the authors is that it cannot provide a
self-consistent time-dependent and non-local convection theory in the usual
sense of these terms.  By looking at the system of equations that define the
theory of convection presented in \cite{2014MNRAS.445.3592P}, see their
eqs. [60]\footnote{Throughout this paper we denote the equations in
  \cite{2014MNRAS.445.3592P} with square brackets to differentiate them from
  our own equations.}, it becomes apparent that their theory is a local
formulation, very much in the spirit of MLT. In a local theory of convection,
velocities and convective fluxes depend only on the local thermodynamical
variables and their local gradients. Usually, a local theory of convection
results from a ``local'' approach to the problem of convection. In a local
approach, one makes the assumption that all the relevant processes are taking
place on length-scales $l$ that are much smaller than the typical length scale
over which the background is changing, i.e. $ l \ll H_p, H_\rho$, where $l$ is
the length scale of the process of interest, $H_p$ and $H_\rho$ are the
pressure and density scale height, respectively. Clearly, the work presented
in \cite{2014MNRAS.445.3592P} follows such a local approach, as clearly stated
in their Sect. 2\footnote{Where they state, \emph{``We proceed further with an
    additional simplification by assuming that the stellar fluid is
    incompressible and irrotational on large distance scales. The concept of a
    large distance scale for incompressibility and irrotationality is defined
    here from a heuristic point of view: This length should be large enough to
    contain a significant number of convective elements so that a statistical
    formulation is possible when describing the mean convective flux of energy
    (see below), but small enough so that the distance traveled by the
    convective element is short compared to the typical distance over which
    significant gradients in temperature, density, pressure, etc. can develop
    (i.e. those gradients are locally small).''} }. It is not possible to
derive a self-consistent non-local theory of convection from such a local
approach, as it is precisely the local approach that decouples the problem at
each
radius.  
Furthermore, a ``time-dependent'' theory of convection has a very specific
meaning in the field of 1D stellar structure computations. It refers to a
theory which is able to describe convection in the case where the stellar
background evolves on a timescale smaller, or of the same order, than the
convective turn-over timescale.  As mentioned in the introduction, such
theories exist but their predictive power is hampered by several free
parameters. As admitted by the authors in one of their footnotes, the theory
presented in \cite{2014MNRAS.445.3592P} is not ``time-dependent'' in the usual
sense\footnote{This is hinted by the authors at the end of section 2,
  p. 3594; \emph{``Before starting our analysis, in order to avoid a possible
    misunderstanding of the real meaning of some of our analytical results, it
    might be wise to call attention to a formal aspect of the mathematical
    notation we have adopted. For some quantities $Q$ function of time or
    space or both, $Q(x;t)$, we look at their asymptotic behaviour by
    formally taking the limits $Q^\infty=\lim_{x\rightarrow x^\infty,\\
      t\rightarrow \infty} Q(x;t)$. This does not mean that we are taking
    temporal intervals infinitely long, rather that we are considering time
    long enough so that the asymptotic trend of the quantity Q is reached but
    still short enough so that the physical properties of the whole system
    have not changed significantly, such as that the star still exists.''}.}. 
  Very likely, \cite{2014MNRAS.445.3592P} refer
to their theory as being ``time-dependent'' because they integrate in time a
set of equations until an asymptotic regime is obtained. However, their theory
of convection is based on the asymptotic regime, where the time variable is
not relevant any more and, consequently it cannot be considered as a theory of
time-dependent convection.

Having clarified that the approach derived by  \cite{2014MNRAS.445.3592P}
deals with a time-independent and local theory we now turn to analyze some of
the mathematical and physical approximations made in their derivation of the
equation of motion for the spherical bubble.

\subsection{The physical assumptions}
\label{sec:physical}
After deriving the equations for the velocity field of an incompressible and
irrotational fluid around an expanding sphere moving within a fluid of
constant density and in hydrostatic equilibrium at infinity (see their
sections 2 and 3), the authors apply this result to compute the forces exerted
on the sphere by the surrounding fluid. Besides the assumptions of an
incompressible and irrotational fluid of constant density, they also neglect
 heat diffusion and restrict themselves to the subsonic regime
(i.e. spheres moving at speeds much smaller than the speed of sound). In this
context they claim that it is reasonable to assume that (see their eq. [12])
\begin{equation}
\frac{v_b}{\dot{R}}\ll 1,\ \forall t>t_{\rm min},
\label{eq:silly_app}
\end{equation}
i.e. that the relative velocity $v_b=|\mathbf{v_b}|$ between the convective element and the intrastellar medium is much smaller than
its expansion velocity $\dot{R}=|\mathbf{\dot{R}}|$--- throughout this work we
denote the radius of the bubble by $R$, and its temporal changes by $\dot{R}$
and $\ddot{R}$. However, it is easy to show that such a regime is in strong
contradiction with the assumptions of a subsonic regime and a local approach
---the latter materialized by the possibility of solving the movement of the
bubble assuming a medium of constant density.

Let us say that a bubble is characterized by its mass $m_b$ (constant in
time), density $\rho_b(t)$, pressure $P_b(t)$, radius $R(t)$, position
$r_b(t)$, and velocity $v_b(t) = \dot r_b$. The surrounding medium is
characterized by its pressure stratification $P(r)$. First, a spherical bubble
traveling in the surrounding medium at a subsonic speed remains in pressure
equilibrium, i.e. $P_b\simeq P$ as sound waves are able to wash out any
pressure difference\footnote{In addition, we show in appendix \ref{app} that
  this is also mathematically consistent with the equations for the dynamics of
  the bubble to be derived later.}. Therefore
$P_b(t) = P(r_b(t))$ and, taking the time derivative, one obtains
\begin{equation}
\frac{\mathrm{d}P_b(t)}{\mathrm{d}t}=\frac{\mathrm{d}P(r_b(t))}{\mathrm{d}t}=\frac{\mathrm{d}P}{\mathrm{d}r}\, v_b =-\frac{Pv_b}{H_P},
\end{equation}
or simply
\begin{equation}
\label{eq:DPeDt}
\frac{\mathrm{d}\log P_b}{\mathrm{d}t}=-\frac{v_b}{H_P}.
\end{equation}
We used the definition of the pressure scale-height $H_P = -
\frac{\mathrm{d} r}{\mathrm{d} \log P}$. Neglecting heat conduction, the
change in density of the bubble follows the adiabatic relation
\begin{equation}
\frac{P_b}{{\rho_b}^{\Gamma_1}}={\rm const},
\end{equation}
\noindent where $\Gamma_1$ is the first adiabatic index. This is equivalent to
\begin{equation}
\label{eq:adiabatic}
\frac{\mathrm{d}\log P_b}{\mathrm{d}t}=\Gamma_1 \frac{\mathrm{d}\log \rho_b}{\mathrm{dt}}.
\end{equation}
Combining eqs. \ref{eq:DPeDt} and \ref{eq:adiabatic}, we obtain:
\begin{equation}
\Gamma_1\, \frac{\mathrm{d} \log \rho_b}{\mathrm{d}t} =-\frac{v_b}{H_P}.
\end{equation}
Finally, as the mass of the bubble is constant, its density decreases as
$\rho_b\propto {R}^{-3}$. Thus,  we obtain:
\begin{equation}
-3\, \Gamma_1 \frac{\mathrm{d} \log R}{\mathrm{d}t} =-\frac{v_b}{H_P}.
\end{equation}
It follows that, within the adiabatic and subsonic approximations, the relation between the expansion rate and the velocity of the bubble is
\begin{equation}
\frac{v_b}{\dot{R}}=\frac{3 H_P\Gamma_1}{R}.
\label{conclusion}
\end{equation}
We can conclude that the assumption $v_b/\dot{R}\ll 1$ is equivalent to $
H_P/R\ll 1$, as usually $\Gamma_1\sim 1$. 

This result can be understood on the basis of the following very simple
physical observation. Within the subsonic approximation, the only way in which
a bubble can expand much faster than it moves is when small vertical
displacements lead to big changes in the pressure of the surrounding fluid,
i.e. when $H_P$ is very small compared to the size of the bubble.

Unfortunately, assuming $v_b/\dot{R}\ll 1$, which implies $ H_P/R\ll 1$, is in
complete contradiction with the core of the theory which is based on a local
picture of convection. In particular it is in clear contradiction with
expressions such as eqs. [3], [13], [24] and [27] from
\cite{2014MNRAS.445.3592P} which are derived within the picture of a bubble
moving in a constant density background.

\subsection{The mathematical approximations}
While the previous inconsistency is serious enough to render the applicability
of the theory questionable, other contradictions develop as a
consequence of mathematical simplifications during the derivation of the
force exerted by the fluid over the moving sphere ---Sections 4.2 and 5 of
\cite{2014MNRAS.445.3592P}. The first of these approximations comes  during
the derivation of ``Lemma 1'' of \cite{2014MNRAS.445.3592P}
(eq. [13]). There it is stated that,
 under the
validity of $v_b/\dot{R}\ll 1$, it is possible to say that
\begin{equation}
\left(\frac{v_b}{\dot{R}}\right)^2 \frac{1}{2}
\left(\frac{9}{4}\sin^2\theta-1\right)\ll
\dot{v}_b\frac{R}{\dot{R}^2}
\left(\frac{3}{2}\cos\theta-\cos\phi\right)
+\frac{\ddot{R}R}{\dot{R}^2},
\label{appro1}
\end{equation}
and also that
\begin{equation}
\left(\frac{v_b\dot{R}}{\dot{R}^2}\right)^2
\frac{5}{2}\cos\theta\ll
\dot{v}_b\frac{R}{\dot{R}^2}
\left(\frac{3}{2}\cos\theta-\cos\phi\right)
+\frac{\ddot{R}R}{\dot{R}^2}.
\label{appro2}
\end{equation}
It is clear that it is not possible to justify these two inequalities
(eqs. [14] in \citealt{2014MNRAS.445.3592P}) solely on the base of
$v_b/\dot{R}\ll 1$ without any other assumption. In order to justify
eqs. \ref{appro1} and \ref{appro2} one must make the assumptions that
$v_b^2\ll |\dot{v}_bR|$ and $v_b^2\ll|\ddot{R}R|$. These two assumptions restrict
even more the physical regime in which the theory could be applicable.  One
might wonder whether such specific regime, i.e.  $v_b/\dot{R}\ll 1$, $v_b^2\ll
|\dot{v}_bR|$ and $v_b^2\ll|\ddot{R}R|$, does exist at all.

We will show later that the two incorrect approximations performed in
eqs. \ref{appro1} and \ref{appro2} do not change the shape of the equation for
the acceleration of the fluid element, although they do change some of the
coefficients. Unfortunately, after the derivation the equation of motion
(their eq. [24])\footnote{We have corrected the sign of the first term,
  because when $M>m_b$ (more buoyancy than weight) the direction of
  $\mathbf{\dot{v}_b}$ should be opposite to that of $\mathbf{g}$ and, also,
  have added the denominator of the first term $(m_b+M/2)$ which should also
  appear in the second term.}
\begin{equation}
\label{eq:acceleration_them}
\mathbf{\dot{v}_b}=\mathbf{g}\frac{m_b-M}{m_b+M/2}-\frac{10}{3}
\frac{\pi {R}^2 \rho \mathbf{v_b} \dot{R}}{m_b+M/2},
\end{equation}
the authors simplify this expression by neglecting the second term to obtain
their eq. [26].  It is not possible to neglect the second term solely on the
base of $v_b/\dot{R}\ll 1$ as it is claimed by \cite{2014MNRAS.445.3592P}.
The physical regime in which this term can be neglected is discussed below.
It is worth noting that their eq. [26] plays a key role in the derivation of
the convective theory, as it is eq. [26] that is used in the further
development of the work ---e.g. in the derivation of their eq. [27].
Interestingly, by doing this the authors dropped the only term that could
provide them with a truly asymptotic regime, as we will show in section
\ref{sol_motion}.  It is easy to see that, the actual physical regime in which
the second term becomes negligible is the one of strong buoyancy forces
$(M-m_b)/m_b\sim 1$.  A simple rewriting of their eq. [24] using the
definition of $M=4\pi R^3 \rho/3$, shows that
\begin{equation}
\label{eq:accel_2}
\mathbf{\dot{v}_b}=\mathbf{g}\frac{m_b-M}{m_b+M/2}-\frac{10}{4}
\frac{M \mathbf{v_b} (\dot{R}/R)}{m_b+M/2}.
\end{equation}
It follows that, for strong buoyancy forces, the second term becomes negligible
when $g\gg v_b \dot{R}/R$. Using that $H_P=P/g\rho$ and that for an ideal gas
the sound speed is ${c_s}^2=\gamma P/\rho$, we see that the second term
becomes negligible if ${c_s}^2/\gamma H_p \gg \dot{R}/R v_b$. As the
derivation of the equation of motion within a local picture requires $H_P\gg
R$, the previous condition holds as soon as ${c_s}^2/\gamma\gg \dot{R}
v_b$. As a result, we see that the second term is indeed negligible as soon as
we have significant buoyancy forces $(M-m_b)/m_b\gtrsim 1$ and we restrain
ourselves to subsonic motions and expansions. The previous argument shows
that, although for very different reasons, in the regime of significant
buoyancy and subsonic bubbles the key equation [26] of
\cite{2014MNRAS.445.3592P} is valid.

Finally, a serious inconsistency  arises during their computation
of the convective flux in their section 6. In order to compute the velocity of
the convective elements (their eq. [41]) the authors analyze the movement of
the stagnation points in the case of a non-expanding rigid-body movement
($\dot{R}=\ddot{R}=0$). The approximation of a non-expanding convective
element is in stark contradiction with the previous derivation of
theory. Furthermore, the authors wrongly assume that $P/\rho+\Phi_g\simeq 0$
at the stagnation points. From this analysis, \cite{2014MNRAS.445.3592P}
conclude that the velocity, radius and acceleration of the bubble are
connected by (see their eq.[41])
\begin{equation}
v_b^2=-\dot{v}_b R.
\label{eq:V2_AR}
\end{equation}
Clearly, assuming eq. \ref{eq:V2_AR} is in apparent contradiction with eqs.
\ref{appro1} and \ref{appro2}, which require $v_b^2\ll|\dot{v}_b R|$. The
neglection of the second term in eq. \ref{eq:accel_2} is also in contradiction
with the simultaneous assumption of $v_b^2=-\dot{v}_b R$ and
eq. \ref{eq:silly_app} ---as these assumptions imply
$v_b\dot{R}/R\gg\dot{v}_b$. We will show in section \ref{sol_motion} that the
ratio $v_b^2/(\dot{v}_b R)$ changes by orders of magnitude during the actual
motion of the bubble (see Figs. \ref{fig:bubble_vb_rb} and
\ref{fig:bubble_vb_rb_2}). Consequently eq. \ref{eq:V2_AR} does not hold.

\section{Equation of motion for an expanding sphere in a potential flow}
\label{eq_motion}
As mentioned during the introduction, during the study of
\cite{2014MNRAS.445.3592P} we found that the equivalent of their key equation
[24] (eq. \ref{eq:acceleration_them}) can be derived in a sound
physical and mathematical way. This is an interesting result which will allow us
to study the motion of an isolated bubble within the present picture and assess
its applicability to derive a theory of stellar convection.

In line with \cite{2014MNRAS.445.3592P} we will assume that the fluid is ideal
(no viscosity), incompressible ($\nabla \cdot\mathbf{v}=0$) and irrotational
($\nabla \times \mathbf{v}=0$). We will assume that the path traveled by the
sphere ($l_b$) can be considered small compared to the distances over which
pressure $P$, gravity $\mathbf g$ or density $\rho$ change.  If $H_P$ and
$H_\rho$ are the pressure and density scale heights we have $l_b\ll H_P$ and
$l_b\ll H_\rho$. The medium is assumed to be in hydrostatic equilibrium far
from the moving element ($\nabla P_\infty=\rho \mathbf{g}$; where $P_\infty$
means the pressure in that layer and far away from the bubble).

\subsection{Flow around an expanding sphere moving at constant velocity}
Under the assumption $\nabla \times \mathbf{v}=0$ there is a potential
$\psi$ so that $\nabla \psi =\mathbf{v}$. The potential of an
incompressible flow of constant density must fulfill $\nabla^2 \psi=0$ --- see
section 9 of \cite{Landau1987} for a detailed discussion of potential
flows. In particular the solution corresponding to the motion (with velocity
$\mathbf{v_b}=v_b\, \mathbf{e_z}$) of an expanding sphere (of radius $R$ and
expansion rate $\dot{R}$) within a fluid which is in hydrostatic equilibrium
far away (i.e. $|\mathbf{x}|\rightarrow \infty$) can be obtained by solving
\begin{equation}
\nabla^2 \psi=0,
\label{eq:solucion}
\end{equation}
with the boundary conditions
\begin{eqnarray}
\label{eq:borde1}
\forall t, \lim_{|\mathbf{x}|\rightarrow \infty} \mathbf{v}&=& 0, \\
\forall t, \forall \mathbf{n'},
\mathbf{v}\cdot \mathbf{n'}&=&\dot{R}+\mathbf{v_b}\cdot \mathbf{n'},\\
&& {\rm  on\ the\ sphere\ } |\mathbf{x}-\mathbf{r_b}|=R,\nonumber
\label{eq:borde2}
\end{eqnarray}
\noindent where we denote the position of the bubble by $\mathbf{r_b}(t)$ and we define $\mathbf{n'}=\mathbf{x'}/|\mathbf{x'}|$, with
$\mathbf{x'}=\mathbf{x}-\mathbf{r_b}$ the position as seen from the center
of the bubble.

It is easier to solve the problem by changing to the coordinate system
comoving with the sphere at constant velocity  $\mathbf{v_b}$. From that
coordinate system the problem reduces to that of an expanding
sphere at rest located at $\mathbf{x'}=0$ within a fluid moving at infinity
with $\mathbf{v_\infty}=-\mathbf{v_b}$, i.e. to solving
\begin{equation}
{\nabla'}^2 \varphi=0,
\end{equation}
where $\nabla'$ denotes the derivatives with respect to $\mathbf{x'}$, with the boundary conditions
\begin{eqnarray}
 \lim_{r'\rightarrow \infty} \mathbf{v'}&=& \mathbf{v_\infty}, \\
\forall \mathbf{n'},
\mathbf{v'}\cdot \mathbf{n'}&=&\dot{R},\ {\rm at\ }\  |\mathbf{x'}|=R,
\end{eqnarray}
\noindent where $\mathbf{v'}={\nabla'\varphi }$ denotes the velocity field as seen from the comoving system, and
$r'=|\mathbf{x'}|$. It is straightforward to check that the solution to that
problem is given by
\begin{equation}
 \varphi(\mathbf{x'}) = \frac{1}{2} \frac{R^3}{{r'}^2} \mathbf v_\infty \cdot
 \mathbf{n'} - \frac{\dot{R} R^2}{r'}  + \mathbf v_\infty \cdot \mathbf{r'}.
\end{equation}
 This is an extension of the solutions
discussed in sections 10 and 11 of \cite{Landau1987} in the case of an
expanding sphere. Computing the derivatives we get
\begin{equation}
\mathbf{v'}= -\frac{3R^3}{2{r'}^3} \mathbf{n'} (\mathbf v_\infty \cdot \mathbf{n'}) + \mathbf{n'} \frac{\dot{R} R^2}{{r'}^2} + \frac{1}{2} \frac{R^3}{{r'}^3} \mathbf v_\infty  +  \mathbf v_\infty.
\label{eq:vel_bubble_rest}
\end{equation}

The velocity field as seen from the system in which the bubble is in
movement with velocity $\mathbf{v_b}$ can be obtained from a direct galilean
transformation:
\begin{equation}
\mathbf{v}{(\mathbf{x})}=\mathbf{v'}{(\mathbf{x'})}+\mathbf{v_b}=
\mathbf{v'}{(\mathbf{x}-\mathbf{r_b})}+\mathbf{v_b}.
\end{equation}
Using $\mathbf{v_\infty}=-\mathbf{v_b}$ we find,

\begin{equation}
\mathbf{v}= \frac{3R^3}{2{r'}^3} \mathbf{n'} (\mathbf v_b \cdot
\mathbf{n'}) + \mathbf{n'} \frac{\dot{R} R^2}{{r'}^2} - \frac{1}{2}
\frac{R^3}{{r'}^3} \mathbf v_b,
\label{eq:field}
\end{equation}
where it is worth noting that $r'=|\mathbf x- \mathbf {r_b}{(t)}|$ and
$\mathbf{n'}=(\mathbf x- \mathbf {r_b}{(t)})/|\mathbf x- \mathbf {r_b}{(t)}| $ are
functions of $\mathbf{x}$ and $t$. One can show that this velocity
fields satisfies  eqs. \ref{eq:solucion}, \ref{eq:borde1} and
\ref{eq:borde2}. This can be easily shown by noting that $\mathbf{x'}=\mathbf{x}-\mathbf{r_b}{(t)}$ implies that for any function
$F(\mathbf{x})$, $\nabla'F(\mathbf{x})=\nabla
F(\mathbf{x})$, $\forall t$. The potential $\psi$ that
produces the field $\mathbf{v}$ (eq. \ref{eq:field}) is given by
\begin{equation}
\psi(\mathbf{x})= -\frac{1}{2} \frac{R^3}{{r'}^2} \mathbf{v_b} \cdot
 \mathbf{n'} - \frac{\dot{R} R^2}{r'}.
\label{eq:potential}
\end{equation}

\subsection{The instantaneous adjustment hypothesis}
In the following, we will assume that the shape of the velocity field instantaneously
adjusts itself to the shape prescribed by eq. \ref{eq:field} for the
instantaneous values of $\mathbf{v_b}(t)$, $R(t)$ and $\dot{R}(t)$, i.e. we
assume that
\begin{equation}
\forall t,\ \mathbf{v}(\mathbf{x},t)= \frac{3{R{(t)}}^3}{2{r'}^3} \mathbf{n'} (\mathbf v_b \cdot
\mathbf{n'}) + \mathbf{n'} \frac{\dot{R}(t) {R{(t)}}^2}{{r'}^2} - \frac{1}{2}
\frac{{R{(t)}}^3}{{r'}^3} \mathbf v_b,
\label{eq:field_t}
\end{equation}
where the position of the bubble is given by $\mathbf{r_b}(t)$ and
$\mathbf{n'}=\mathbf{x'}/|\mathbf{x'}|$, where
$\mathbf{x'}=\mathbf{x}-\mathbf{{r_b}{(t)}}$ is the position as seen from the
center of the bubble. The velocity field of eq. \ref{eq:field_t} fulfills the
boundary conditions given by eqs. \ref{eq:borde1} and \ref{eq:borde2} at every
time $t$. As $t$ and $\mathbf{x}$ are independent variables, it is easy to
show that the potential $\psi(\mathbf{x},t)$ that produces this field is
\begin{equation}
\psi(\mathbf{x},t)= -\frac{1}{2} \frac{{R{(t)}}^3}{{r'}^2} \mathbf{v_b} \cdot
 \mathbf{n'} - \frac{\dot{R}(t) {R(t)}^2}{r'}.
\label{eq:potential_t}
\end{equation}

In order for this hypothesis to hold, the fluid needs to
adjust fast enough to the instantaneous velocity of the bubble. This
hypothesis will hold if both the expansion velocity of the sphere and the
translational velocity of the sphere are much smaller than the sound speed,
i.e. if $v_b\ll c_s$ and $\dot{R}\ll c_s$. In addition we also assume that the
timescales related to the acceleration and the change in the expansion rate
are small compared with the reaction timescale of the fluid given by
$\tau=R/c_s$--- i.e. we assume that changes in $v_b$ and $\dot{R}$ fulfill
$\dot{R}/\ddot{R}\ll R/c_s$ and $v_b/\dot{v}_b\ll R/c_s$. Under the assumption
of subsonic flows, this implies that $|\dot{\mathbf{v}}_b|\ll c_s^2/R $ and
$\ddot R\ll c_s^2/R $.  Note that the assumption of subsonic velocities is
also compatible with the incompressibility approximation, which implies $
c_s=\infty$.

\subsection{Equation of motion for a moving and expanding sphere within a
  fluid at rest}
Once the velocity field is known, one can use this result to compute the
force exerted by the fluid on the moving bubble by using
Euler's equation
\begin{equation}
 \partial_t \mathbf v + (\mathbf v \cdot \mathbf \nabla) \mathbf v = - \frac{\mathbf \nabla P}{\rho} + \mathbf g,
\label{eq:motion_a}
\end{equation}
where $\mathbf g = - \mathbf \nabla \Phi_g$ is the gravitational acceleration.
For an incompressible and irrotational fluid of constant
density, eq. \ref{eq:motion_a} be written as
\begin{equation}
 \partial_t \mathbf v + \mathbf \nabla \frac{\mathbf v^2}{2} = - \mathbf \nabla \frac{P}{\rho} - \mathbf \nabla \Phi_g.
\label{eq:motion2}
\end{equation}
Eq. \ref{eq:motion2} can then be rewritten, using  $\nabla \psi =\mathbf{v}$, as
\begin{equation}
\nabla\left ( \partial_t \psi +  \frac{\mathbf v^2}{2} +
  \frac{P}{\rho} +  \Phi_g\right )=0.
\label{eq:Bernoulli}
\end{equation}
Integrating this equation in space we find
\begin{equation}
\partial_t \psi +  \frac{\mathbf v^2}{2} +  \frac{P}{\rho} +  \Phi_g +c(t)=0,
\label{eq:Bernoulli2}
\end{equation}
where $c(t)$ is a constant of integration.  It can be obtained by noting that
for $|\mathbf{x}|\rightarrow \infty$ the fluid is static ($\mathrm{v}=0$) and
in hydrostatic equilibrium ($\nabla( P/\rho + \Phi_g)=0$). This implies
that\footnote{Note that here the expression $|\mathbf{x}|\rightarrow \infty$
  means in fact at $|\mathbf{x}-\mathbf{r_b}|\gg R$. Strictly speaking the
  limit $|\mathbf{x}|\rightarrow \infty$ is ill-defined for a gravitational
  potential of a constant gravity field. Also, note that, as we are assuming
  that the hydrostatic pressure changes in much larger distances we are
  considering that at $|\mathbf{x}-\mathbf{r_b}|\gg R$ the pressure $P_\infty$
  depends on $z$ so that it can balance the changes in $\Phi_g(z)$. Due that
  at the scales of the problem $P_\infty$ remains almost constant, also
  $\Phi_g$ must remain almost constant. In this context it is useful to think
  the limit $|\mathbf{x}|\rightarrow \infty$ on the $xy$-plane, where $\Phi_g$
  and $P_\infty$ are in fact strictly constant. Then the choice of $C'=0$
  corresponds to choosing $\Phi_g=-g (z-r_b)-P_\infty(z=r_b)/\rho$}
\begin{equation}
\left( \frac{P}{\rho} + \Phi_g\right)_{|\mathbf{x}|\rightarrow
\infty}=C',
\end{equation}
where $C'$ is a constant that depends on the arbitrary choice of the
definition of the gravitational potential.  Noting that for
$|\mathbf{x}|\rightarrow \infty$ we have that $ \partial_t \psi \rightarrow
0$ and $v^2 \rightarrow 0$, we see that eq. \ref{eq:Bernoulli2} implies that
$c(t)=-C'$. For the sake of simplicity we can set $C'=c(t)=0$, and we obtain
\begin{equation}
  \frac{P}{\rho}= - \partial_t \psi - \frac{\mathbf v^2}{2} -  \Phi_g.
\label{eq:presion}
\end{equation}

The force $\mathbf{F}$ applied to the bubble is obtained by
integrating eq. \ref{eq:presion} over the surface of the sphere
$\partial V{(t)}$,
\begin{align}
 \mathbf{F}=&-\int_{\partial V} P \mathbf{n'} dS \\
 =& \rho \int_{\partial
   V} \partial_t \psi  \mathbf{n'} dS+\rho \int_{\partial V} \frac{\mathbf v^2}{2}  \mathbf{n'} dS+\rho \int_{\partial V} \Phi_g \mathbf{n'} dS.
\label{eq:force}
\end{align}

The first integral in the RHS of eq. \ref{eq:force} can be obtained using the
definition of $\psi$, taking the time derivative $\partial_t \psi$ and
evaluating over the sphere. We have
\begin{equation}
\partial_t \psi=-\frac{3}{2} \dot{R} (\mathbf{v_b}\cdot \mathbf{n'})
-\frac{R}{2}  (\mathbf{\dot{v}_b}\cdot \mathbf{n'})
-\ddot{R} R - 2 \dot{R}^2,\ \ {\rm for}\ \ |\mathbf{x-r_b(t)}|=R.
\end{equation}
Integrating over the whole sphere we get
\begin{equation}
\int_{\partial V} (-\partial_t \psi) \mathbf{n'}  dS=
2\pi R^2 \dot{R} \mathbf{v_b} +\frac{2\pi}{3} R^3 \mathbf{\dot{v}_b}.
\label{eq:dpsi}
\end{equation}

The second integral in the RHS of eq. \ref{eq:force} can be directly computed
once the velocity field is evaluated over the surface of the sphere:
\begin{equation}
\mathbf{v}{(\mathbf{x})}=\left(
v_b \cos \theta + \dot{R}\right) \mathbf{n'} +
v_b \frac{\sin \theta}{2} \mathbf{e_\theta},\ \ {\rm at}\ |\mathbf{x-r_b(t)}|=R(t),
\label{eq:v_esfera}
\end{equation}
where we have defined the spherical coordinates $r'$, $\theta$ (zenithal angle)
and $\phi$ (azimuthal angle) measured from the instantaneous center of the
sphere, and $\mathbf{e_\theta}$ is the unitary vector in the azimuthal
direction. From eq. \ref{eq:v_esfera} we get
\begin{equation}
{v{(\mathbf{x})}}^2=\left(
v_b \cos \theta + \dot{R}\right)^2 +
\left(v_b \frac{\sin \theta}{2}\right)^2,\ \ {\rm at}\ |\mathbf{x-r_b(t)}|=R(t).
\label{eq:v_esfera2}
\end{equation}
Integrating over the whole sphere we get
\begin{equation}
\int_{\partial V} \frac{v^2}{2} \mathbf{n'} dS=\frac{4\pi}{3} \dot{R} R^2\mathbf{v_b},
\end{equation}
where we have used that $\mathbf{e_z}=\mathbf{e_{z'}}$ and $\mathbf{v_b}=v_b\,
\mathbf{e_z}$.

Finally, the last integral in the RHS of eq. \ref{eq:force} can be integrated using that
\begin{equation}
\int_{\partial V} \Phi_g \mathbf{n'} dS = \int_{V} \nabla\Phi_g dV =
-\mathbf{g} V(t),
\label{eq:gravity}
\end{equation}
where $V(t)=4\pi R(t)^3/3$ is the volume of the expanding sphere.

Using eqs. \ref{eq:dpsi}, \ref{eq:v_esfera2}, and \ref{eq:gravity} in
eq. \ref{eq:force}, the force exerted by the fluid on the moving
bubble is
\begin{equation}
 \mathbf{F}=-\int_{\partial V} P \mathbf{n'} dS =
-\frac{4\pi R^3}{3}\rho \mathbf{g}-\frac{2\pi}{3}\rho R^2 \dot{R} \mathbf{v_b}
-\frac{2\pi}{3}\rho R^3 \mathbf{\dot{v}_b}.
\label{eq:force_final}
\end{equation}

\subsection{The acceleration of the bubble}
The equation of motion for the moving sphere, under all the previously
mentioned assumptions, is
\begin{equation}
m_b \mathbf{\dot{v}_b}=-\int_{\partial V} P \mathbf{n'} dS + m_b
\mathbf{g},
\label{eq:motion_b}
\end{equation}
where $m_b$ is the mass of the bubble ($m_b=4\pi R^3 \rho_b/3$), and the pressure integral is given by
\ref{eq:force_final}. Using the definition $M=4\pi R^3 \rho/3$ (i.e. the mass
of a bubble of same radius but with the density of the fluid)
eq. \ref{eq:motion_b} gives a very simple expression for the acceleration of the
bubble;
\begin{equation}
\mathbf{\dot{v}_b}= \frac{(m_b-M)}{(m_b+M/2)}\mathbf{g} -
\frac{M}{2(m_b+M/2)} \frac{\dot{R}}{R} \mathbf{v_b}.
\label{eq:acceleration}
\end{equation}
This is the correct version of the acceleration derived by
\cite{2014MNRAS.445.3592P} in their eq. [24]. The first thing that is apparent
from the first term in eq. \ref{eq:acceleration} is that, in the regime
corresponding to our physical approximations, the acceleration of a bubble at
rest is smaller by a factor $1+M/(2\, m_b)$ compared with the Archimedes
principle for a static fluid. While this might be surprising at
first glance, its physical explanation is quite simple. Within the
approximation of eq. \ref{eq:field_t} the fluid is forced to be accelerated
when the bubble is accelerated. By looking at the stagnation points on top and
below the bubble it becomes clear that the fluid there moves at every time at
the same velocity as the bubble. In order to fulfill Euler's equation
for a velocity field that changes with time some forces must be exerted at the
boundary of the fluid (and equivalently, its reaction felt on the moving
bubble). Consequently, the factor $1+M/(2\, m_b)$ accounts for the fact that,
in order to accelerate, and fulfill eq.  \ref{eq:field_t}, the bubble must
carry the nearby fluid with it. The force exerted on the bubble by the
surrounding medium is also responsible for the second term in
eq. \ref{eq:acceleration}. In this case the term arises from the fact that, as
the bubble expands, more fluid needs to be accelerated to fulfill eq.
\ref{eq:field_t}. This term acts in the same orientation as the velocity, but
its direction is determined by the sign of $\dot{R}$. Depending on
whether the bubble is expanding or contracting, this term acts in the same
direction as the velocity $\mathbf{v_b}$ or in the opposite one. In the latter case,
it acts as a drag. It is worth noting that the claim of \cite{2014MNRAS.445.3592P} that this drag-like
term reconciles the potential flow approximation with d'Alembert paradox is
wrong, as this force is only present in the case of contracting or expanding spheres,
and it is in no way related to real drag forces, which can be of viscous
or turbulent origin. This is apparent from the fact that the force acts
in the opposite direction, than that of a real drag force, in the case of
contracting bubbles. Also, it is easy to see from eq. \ref{eq:vel_bubble_rest}
that the relative velocity of the fluid and the sphere has a tangential
component at the surface of the sphere, contrary to what is known to happen at
boundary layers.

 Eq. \ref{eq:acceleration} has been derived under the assumption
  that the flow remains irrotational (potential) at all times. This is
  a very strong physical assumption and it would be necessary to
  investigate to which extent this will be an appropriate description
  of a given real fluid. For a compressible, viscous fluid moving
  under a conservative body force, we have that the vorticity
  ($\nabla\times\mathbf{v}$) fulfills
\begin{eqnarray}
\frac{\mathrm{D}(\nabla\times\mathbf{v})}{\mathrm{D}t}=\left((\nabla\times\mathbf{v})\cdot\nabla\right)\mathbf{v}&-&(\nabla\times\mathbf{v})(\nabla\cdot\mathbf{v})\nonumber\\
+\nabla\times\left(\frac{\nabla\cdot\tau}{\rho}\right)&+&\frac{\nabla \rho\times \nabla P}{\rho^2},
\label{eq:vorticity}
\end{eqnarray}
where $\mathrm{D}/\mathrm{D}t$ denotes the Lagrangian derivative and
$\tau$ is the viscous stress tensor. In the general case, density
will depend both on temperature and pressure. This implies that, in
most cases $\nabla \rho\times \nabla P\neq 0$.  Even if the flow is
irrotational at the beginning of motion, one should expect that
vorticity ($\nabla \times\mathbf{v}$) will be created at later times
in a real flow by the last term in the RHS of
eq. \ref{eq:vorticity}. In addition, the absence of a drag force in
eq. \ref{eq:acceleration} reminds us of the existence of boundary layers
in real fluids around solid bodies, where viscosity cannot be
completely neglected. In boundary layers, the third term in the RHS of
eq. \ref{eq:vorticity} will also lead to the creation of
vorticity. Consequently, even if the initial condition is that of an
irrotational flow, there is no reason to expect that the flow will
remain irrotational at all times. Besides the hypotheses done on the
flow, the derivation of eq. \ref{eq:acceleration} also assumes that
the bubble remains spherical at all times. However,
eq. \ref{eq:presion} shows that pressure differences at the surface of
the bubble should deform it as soon as it starts to move, unless
internal forces prevent it (e.g. in a solid body). Because of all
these assumptions, the use of eq. \ref{eq:acceleration} to describe the
movement of spherical bubbles in stellar interiors might not be valid
unless proven otherwise for each particular case.

Finally, up to now we have not made any assumption on the properties of
the ``bubble" element. However, in a convection theory we want the
bubble to be made of the same material as the surrounding fluid. In
the next section we adopt an equation of state for the fluid inside
the sphere and use it to describe the dynamics of the bubble.

\section{Motion of an isolated bubble  -- solutions and asymptotic behaviors}
\label{sol_motion}
\subsection{General case}
While it is not our aim in this paper to develop a convection theory, we want
to assess the expected behavior for the motion of the bubble under the
equation of motion derived in the previous section. The projected equation of
motion of the bubble in the radial direction is
 \begin{equation}
 \label{eq:motion}
 \dot{v}_b = - \frac{m_b - M}{m_b + \frac{M}{2}}g - \frac{1}{2} \frac{\dot{R}}{R} \frac{M}{m_b + \frac{M}{2}} v_b,
 \end{equation}
 \noindent with $M = 4\pi/3 R^3 \rho$ the buoyant mass, and $m_b$ the bubble
 mass.

 To solve the bubble motion through the whole convective region we apply
 eq. \ref{eq:motion} at a \emph{given location} of a stellar
 stratification. This is the spirit of solving a problem using local approach:
 the force balance that determines the acceleration of the bubble is computed
 in a local approach, and the result is used to determine the motion of the
 bubble through the convective region. This means that we need to specify the
 value of the thermodynamic variables, $T$, $\rho$ and $P$, as well
 as their stratification given by $H_P$, $H_\rho$ and
 $\nabla=\mathrm{d}\log T/\mathrm{d}\log P$. Only four of them can
 be independently set, as they are related by the equation of state
 $\rho(T,P)$ of the stellar material, which implies
\begin{equation}
\frac{\mathrm{d}
  \rho}{\rho}=\alpha\frac{\mathrm{d}P}{P}-\delta\frac{\mathrm{d}T}{T},
\label{eq:EOS}
\end{equation}
and consequently
\begin{equation}
\nabla=\frac{\alpha}{\delta}-\frac{1}{\delta}\frac{H_P}{H_\rho},
\end{equation}
where $\alpha=(\partial \log\rho/\partial \log P)_T$ and $\delta=-(\partial
\log\rho/\partial\log T)_P$. In order to solve eq. \ref{eq:motion} we
need to know the evolution of $R$ and $M$ as the bubble evolves.

The evolution of the buoyant mass $M$ can be easily obtained by taking the time
derivative of its definition:
\begin{equation}
\frac{\dot{M}}{M}=3\frac{\dot{R}}{R}+\frac{\dot{\rho}}{\rho},
\end{equation}
since $\dot{\rho}=\mathrm{d}\rho(r(t))/\mathrm{d}t=-\rho\,
v_b/H_\rho$, we have
\begin{equation}
\frac{\dot{M}}{M}=3\frac{\dot{R}}{R}-\frac{v_b}{H_\rho}.
\label{eq:dotM}
\end{equation}
The evolution of the radius $R$ of the bubble can be obtained from the
equation of state (eq. \ref{eq:EOS}) and the assumption of subsonic motions.
 From eq. \ref{eq:EOS} it is immediate that
\begin{equation}
\frac{\dot{\rho}_b}{\rho_b}=\alpha\frac{\dot{P}}{P}-\delta \frac{\dot{T}_b}{T_b},
\end{equation}
where we label with $b$ the thermodynamic quantities inside the bubble, and we
have used that $P_b=P(r(t))$.  Using the fact that the mass of the bubble is constant, i.e. $\dot{\rho}_b/\rho_b=-3\dot{R}/R$, and using eq. \ref{eq:DPeDt}, we finally get that the expansion of the bubble is governed by
\begin{equation}
\frac{\dot{R}}{R}=\frac{\delta}{3}\frac{\dot{T}_b}{T_b}+\frac{\alpha}{3}\frac{v_b}{H_P}.
\label{eq:dotR}
\end{equation}
To solve the dynamics it is still necessary to know the evolution of the
temperature of the bubble $T_b$. This cannot be derived without taking into
account the amount of heat lost (or gained) by the bubble as it moves. The
energy balance of the bubble is given by (see \citealt{2012sse..book.....K}),
\begin{equation}
\frac{\mathrm{d}q}{\mathrm{d}t}=c_P \frac{\mathrm{d}T}{\mathrm{d}t}
-\frac{\delta}{\rho}\frac{\mathrm{d}P}{\mathrm{d}t}.
\label{eq:firstlaw}
\end{equation}
The heat flux $\mathbf{F}$ from the bubble is given by
\begin{equation}
\mathbf{F}=-k_{\rm rad}\mathbf{\nabla}T,\ \hbox{\rm where}\
k_{\rm rad}=\frac{4 a c}{3}\frac{{T_b}^3}{\kappa_b \rho_b}.
\end{equation}
Estimating that the temperature gradient between the bubble and the
surrounding fluid is $\mathrm{d}T/\mathrm{d}R\simeq (T(r)-T_b)/R$,
the heat losses from the spherical bubble are given by
\begin{equation}
\frac{dq}{dt}\simeq\frac{3}{\rho_b R^2}k_{\rm rad} (T(r)-T_b).
\label{eq:dqdt}
\end{equation}
Replacing eq. \ref{eq:dqdt} in eq. \ref{eq:firstlaw} gives
\begin{equation}
\frac{\dot{T}_b}{T_b}\simeq \frac{3 k_{\rm rad}}{\rho_b R^2 c_P}
\left[\frac{T(r)}{T_b}-1\right]-\nabla_{\rm ad}\frac{v_b}{H_P},
\label{eq:dotT}
\end{equation}
where in the second term of the right hand side we replaced $\dot{P}=-P\,
v_b/H_P$, and used that $\nabla_{\rm ad}=(P\delta)/(c_P \rho_b T_b)$.

Eqs \ref{eq:motion}, \ref{eq:dotM}, \ref{eq:dotR} and \ref{eq:dotT},
together with the stratification of the star $P(r)$, $\rho(r)$,
$T(r)$, $H_P(r)$ $H_\rho(r)$ and $\nabla(r)$, allow to solve the
motion of the bubble.  The reader should also be aware, however,
  that in order to use eq. \ref{eq:motion} to describe the motion of a
  bubble in a real flow, one should first show that the flow remains
  irrotational at all times. This is not trivial and in principle
  there is no reason to state that the generation of vorticity will be
  small. Eqs \ref{eq:motion}, \ref{eq:dotM}, \ref{eq:dotR} and
\ref{eq:dotT}, show that, even within the picture developed by
\cite{2014MNRAS.445.3592P}, it is necessary to  take into account
the radiative heat losses from the bubble (eq. \ref{eq:dotT}) before
being able to solve the dynamics of the bubble.  Eq. \ref{eq:dotT}
shows that depending on the typical timescales for the expansion
($\tau_{\rm exp}=H_P/v_b$) and thermal diffusion
($\tau_{th}=\rho_bc_PR^2/3 k_{\rm rad}$) the evolution of $T_b$ will
be completely different. In particular, as $\tau_{th}\propto R^2$,
thermal diffusion always dominates the dynamics for bubbles that are
small enough. In the extreme case in which heat diffusion dominates,
the bubble expands in isothermal equilibrium and there is no buoyancy.
This is in stark contrast with the derivations performed by
\cite{2014MNRAS.445.3592P} who solve (in their sections 4 and 5) the
dynamics of the bubble without taking into consideration the role of
heat diffusion\footnote{In fact, the authors claim at the beginning of
  section 4.2 that the dynamics of the bubble is solved under the
  assumption of adiabatic expansion. However, a careful examination of
  the derivations sheds that this hypothesis is never used.}. It is
only in their section 6, \emph{after} having solved the dynamics of
the bubble, that they consider heat losses from the bubble. We will
show in the next section that solving the dynamics without addressing
the heat lost by the bubble can lead to extremely unphysical results.

In the bulk of the solar convective zone, one has $\tau_{\rm exp}\sim
10^5...10^6$~s and $\tau_{\rm th}\sim 10^{12}\times\eta^2$~s for
convective elements of size $R\sim\eta H_P$ (see
Fig. \ref{fig:sol}). The motion of convective elements in those cases
is very close to adiabatic down to very small sizes
---i.e. $\eta\gtrsim 10^{-3}$. Even in the very outer regions of the
sun, one finds that the expansion timescale is shorter than the
thermal timescales, and the movement of a bubble is close to adiabatic
for convective elements of size $R\sim H_P$. For example, in the
standard solar model of Fig. \ref{fig:sol}
\citep{2008Ap&SS.316...99W}, we see that at $r\simeq 0.999 R_\odot$
one still finds that $\tau_{\rm exp}\sim 10^3$~s and $\tau_{\rm
  th}\sim 10^{6}\times\eta^2$~s and convective elements move almost
adiabatically. While the assumption of adiabaticity is good to study
the motion of convective elements in most of the solar convective
zone, one should keep in mind that it is in the regions far from
adiabaticity that a better convection theory than MLT is needed to
predict the correct value of the temperature gradient $\nabla$.

\subsection{Solutions for the adiabatic motion of the bubble}
It is well known that in the inner convective regions of stars the movement
of convective elements of reasonable size is almost adiabatic due to the high
density of the stellar matter. The assumption of adiabatic expansion greatly
simplifies the treatment of eqs. \ref{eq:motion}, \ref{eq:dotM}, \ref{eq:dotR}
and \ref{eq:dotT}. This allows for an easy test case for the dynamics of the
bubble predicted by the method of \cite{2014MNRAS.445.3592P}.  For the sake of
clarity we will now consider the case of an ideal gas
($\alpha=\delta=1$) with a constant adiabatic index
$\gamma=\Gamma_1=(1-\nabla_{\rm ad})^{-1}=5/3$. In the case of a bubble moving
adiabatically in the stellar medium ($k_{\rm rad}= 0$) eq. \ref{eq:dotT} can
be directly substituted into eq. \ref{eq:dotR} to give
\begin{equation}
\frac{\dot{R}}{R}=\frac{v_b}{3H_P}[1-\nabla_{\rm ad}]=\frac{v_b}{3\gamma H_P}.
\label{eq:dotR-ad}
\end{equation}
Using eq. \ref{eq:dotR-ad} in eq. \ref{eq:dotM} we can derive that
\begin{equation}
\frac{\dot{M}}{M}=\frac{v_b}{H_P}[\nabla-\nabla_{\rm ad}],
\label{eq:dotM-ad}
\end{equation}
where we have used the fact that $\nabla=1-H_P/H_\rho$. The evolution of the
bubble in the adiabatic case is given by the set of equations \ref{eq:motion},
\ref{eq:dotR-ad} and \ref{eq:dotM-ad}. Note that eq. \ref{eq:dotM-ad}
describes the usual Schwarzschild criterion. $M$ is the mass of the fluid that
occupies the same volume as the bubble. If $M>m_b$ the bubble will rise due to
buoyancy, and if $M<m_b$ the bubble will sink due to its own weight. Let us
consider a bubble in equilibrium, i.e. $M=m_b$, but under different values of
$\Delta \nabla=\nabla-\nabla_{\rm ad}$.  When $\Delta \nabla > 0$, a positive
velocity perturbation will lead to an increase in $M$, leading to an upward
force ($M>m_b$). On the other hand, a negative velocity perturbation will lead
to a decrease of $M$ which will lead to a downward force ($M<m_b$). As
expected, an unstable situation results. Similarly, $\Delta \nabla < 0$
($\nabla < \nabla_\mathrm{ad}$) corresponds to a stable situation.

Substituting eq. \ref{eq:dotR-ad} in eq. \ref{eq:motion}, we obtain the final
set of equations that we need to solve:
\begin{eqnarray}
\label{eq:vdot_full}
 \dot{v}_b &=& - \frac{m_b - M}{m_b + \frac{M}{2}}g - \frac{1}{6 \gamma H_p} \frac{M}{m_b + \frac{M}{2}} v_b^2, \\
 \frac{\dot{M}}{M} &=&  \frac{v_b}{H_p} \Delta \nabla.
\end{eqnarray}
It is best to formulate the system using non-dimensional
quantities. We choose to normalize lengths with the pressure scale-height
$H_p$ ($= P/\rho g$), velocities with the sound speed $c_s$ ($= \sqrt{\gamma
  P/\rho}$), and masses with the bubble mass $m_b$. In these units, time is
measured in units of $H_p/c_s$. The normalized system is
\begin{eqnarray}
 \dot{v}_b &=& - \frac{1}{\gamma} \frac{1 - \omega}{1 + \frac{\omega}{2}} - \frac{1}{6\gamma} \frac{\omega}{1 + \frac{\omega}{2}} v_b^2, \label{eq1}\\
 \frac{\dot{\omega}}{\omega} &=& v_b \Delta \nabla. \label{eq2}
\end{eqnarray}
where $\omega = M/m_b$. Writing $m_b = 4\pi/3 R^3 \rho_b$, with $\rho_b$ the
bubble density, one has
\begin{equation}
  \omega(t) = \frac{\rho(r_b(t))}{\rho_b(t)}.
\end{equation}
$\omega$ is the ratio between the background density and the bubble
density. We define the density perturbation of the bubble as $\delta \rho =
\rho_b - \rho$, so that $\delta \rho / \rho = 1/\omega -1$.

The system requires two initial conditions. The first initial condition is the
initial velocity, $v_b(t=0)$; the second initial condition is given by
$\omega(t=0) = \rho(r_b(t=0))/\rho_b(t=0)$, the initial density perturbation
of the bubble. Having normalized lengths to the value of $H_P$ the problem
depends on one other parameter, the superadiabaticity $\Delta \nabla$.

It is worth noting that the radius of the bubble does not enter the adiabatic
motion problem directly. However, once a solution $(v_b(t), \omega(t))$ is
known, the expansion of the bubble can be computed by integrating
eq. \ref{eq:dotR-ad}. In normalized form it writes
\begin{equation}
 \label{eq:bubble_R}
 \frac{\dot R}{R} = \frac{v_b}{3\gamma}.
\end{equation}
We now rewrite it as
\begin{equation}
  \frac{\mathrm{d}}{\mathrm{d} t} \ln R = \frac{\dot{r}_b}{3\gamma},
\end{equation}
which immediately leads to
\begin{equation}
  \ln \frac{R}{R_0} = \frac{r_b}{3 \gamma},
\end{equation}
where $R_0$ is the bubble initial radius. The change in the bubble radius is
directly related to the distance it has traveled from its initial position.

Eqs. \ref{eq1} and \ref{eq2} are solved numerically.  As initial conditions,
we consider that the bubble is at rest, $v_b(t=0) = 0$, and we use a density
perturbation to initiate the motion of the bubble. We explore positive and
negative initial density perturbations of different magnitudes, namely:
$\delta \rho/\rho= -10^{-6}$, $-10^{-3}$, $-10^{-1}$, $-0.5$, $10^{-6}$,
$10^{-3}$, $10^{-1}$, $0.5$ ---note that each $\delta \rho/\rho$ implies a
different $\delta T_b/T_b$ so that pressure is balanced. We also investigate
different values of the superadiabaticity, namely $\Delta \nabla = 10^{-3}$,
$10^{-1}$. These values cover a range going from a nearly adiabatic
stratification, as found in the deep stellar interior, to a value corresponding
to a slight superadiabaticity, as found close to the stellar surface where the
movement of the bubble can still be solved within the assumption of adiabatic
expansion.

\begin{figure*}[t] 
  \centering
  \includegraphics[width=\linewidth]{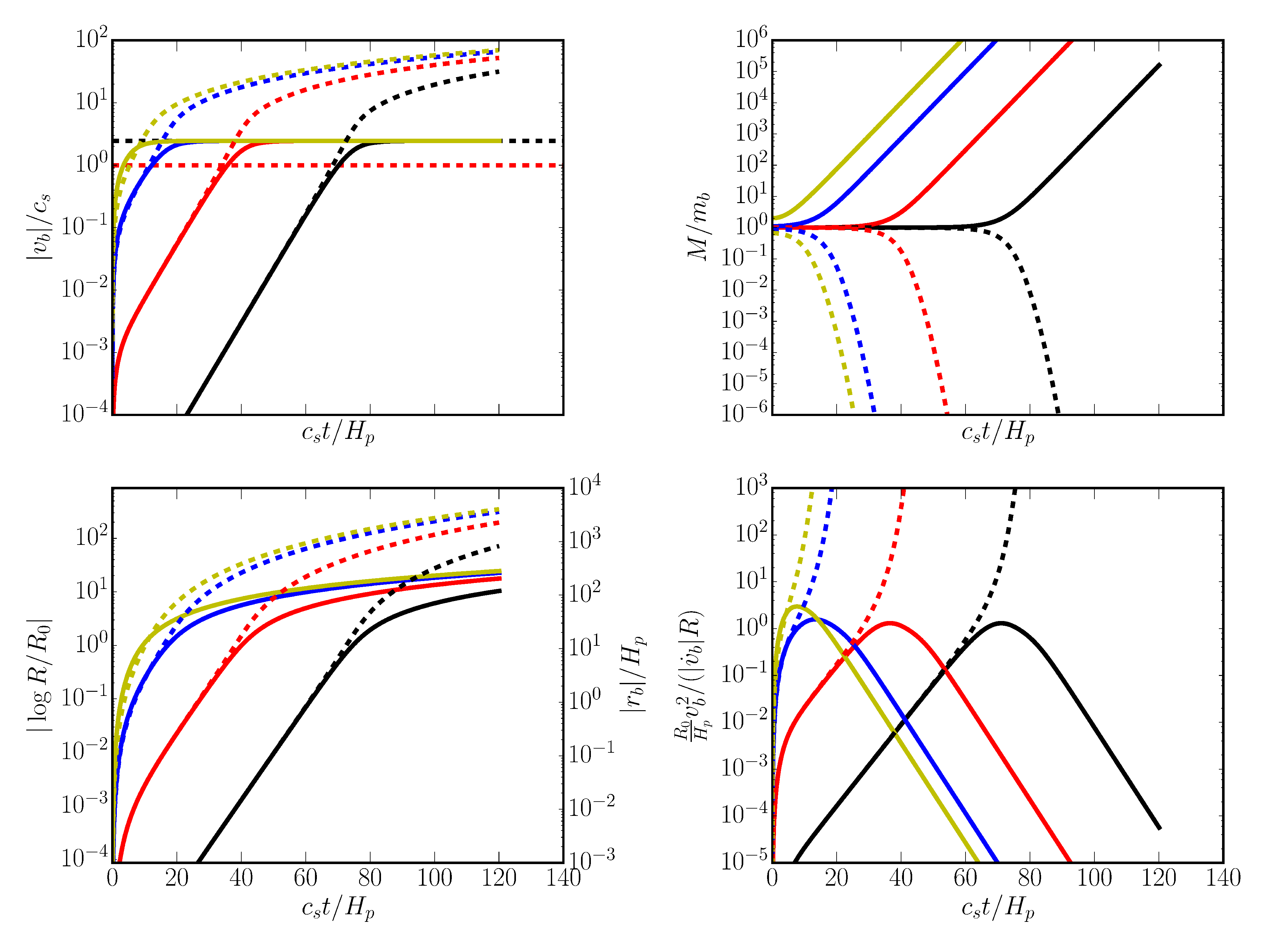}
  \caption{Solution of the bubble motion for $\Delta \nabla = 10^{-1}$ and
    different initial bubble density perturbations (different colors
    correspond to different magnitude of the perturbation). Upper left panel:
    evolution of the bubble's velocity ($|v_b(t)|$). Upper right panel:
    evolution of $\omega=M/m_b$. Bottom left panel: evolution of the bubble's
    expansion and position. Bottom right panel: evolution of the ratio
    ${v_b}^2/(\dot{v}_b R)$. The dashed lines correspond to the cases where
    $\delta \rho>0$, for which $v_b<0$, $r_b<0$, and $\log R/R_0<0$. The red
    horizontal dashed line in the left upper panel shows the (conservative)
    limit $v_b = c_s$ above which the theory is not valid. The black
    horizontal dashed line in the left panels is the asymptotic velocity
    $v_b^\infty = \sqrt{6} \sim 2.45$ (see text).}
  \label{fig:bubble_vb_rb}
\end{figure*}
\begin{figure*}[t] 
  \centering
   \includegraphics[width=\linewidth]{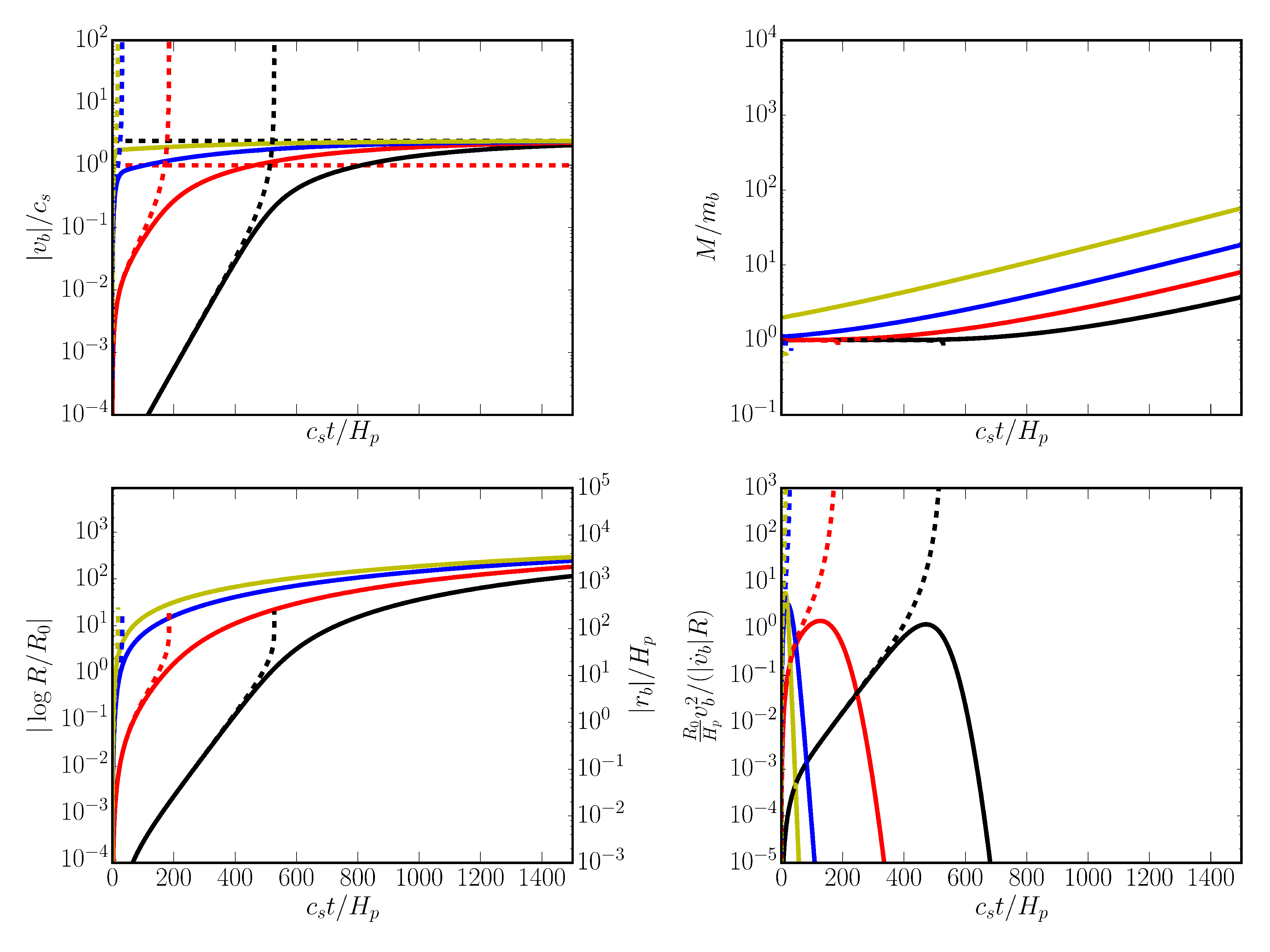}
   \caption{Same as Fig. \ref{fig:bubble_vb_rb} but for the case of $\Delta
     \nabla = 10^{-3}$}
   \label{fig:bubble_vb_rb_2}
\end{figure*}

\begin{figure} 
  \centering
  \includegraphics[width=\linewidth]{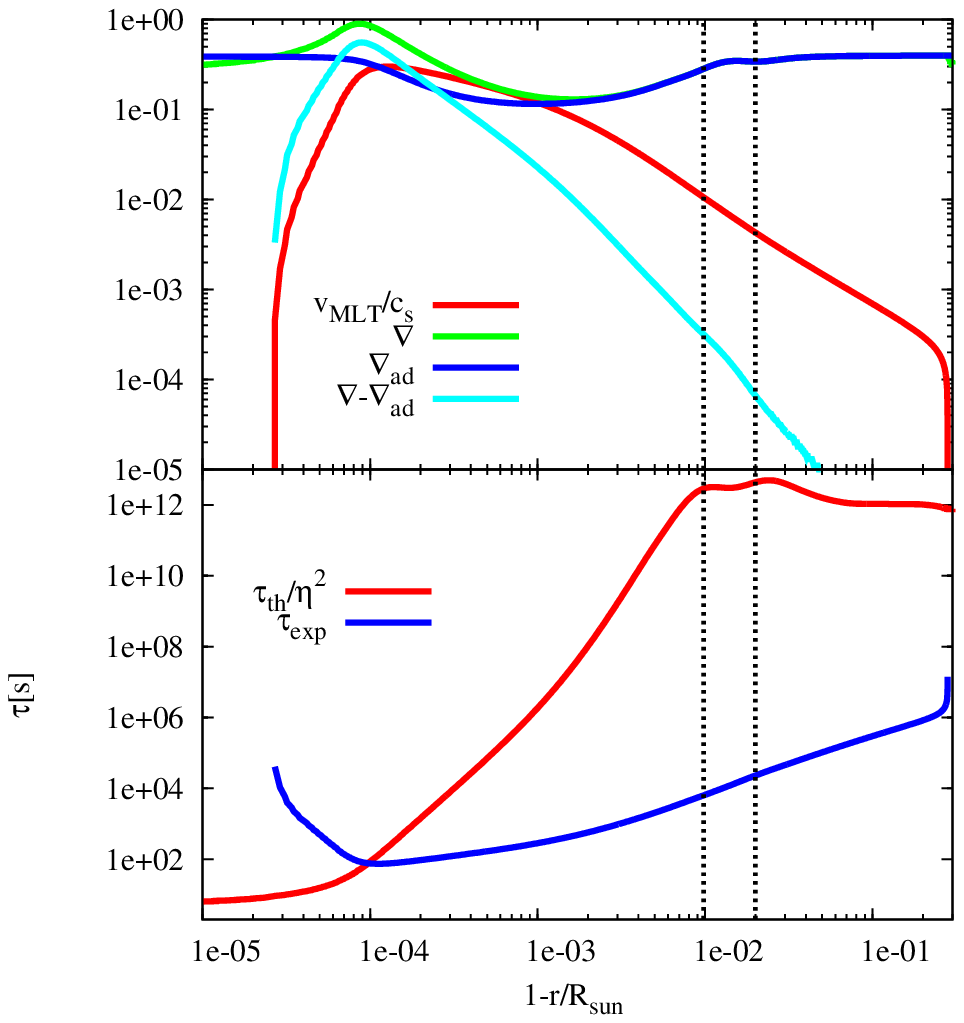}
  \caption{ Properties of convective zone of the Garching solar model
    (GARSOM) as presented in \protect\cite{2008Ap&SS.316...99W}, using the
    \protect\cite{1998SSRv...85..161G} solar composition, and updated nuclear
    reaction rates \protect\citep{2011RvMP...83..195A} and low-temperature
    Rosseland mean opacities \protect\citep{2005ApJ...623..585F}. Upper panel:
    Depth dependence of the convective velocity $v_{\rm MLT}$, actual
    temperature gradient $\nabla$, adiabatic gradient $\nabla_{\rm
      ad}$ and superadiabaticity $\Delta\nabla=\nabla-\nabla_{\rm
      ad}$. Bottom panel: Depth dependence of the thermal diffusion
    timescale $\tau_{\rm th}$ and the expansion timescale $\tau_{\rm
      exp}$ for convective elements of size $R=\eta H_P$. The dotted
    vertical lines denote the layers at $r=0.98 R_\odot$ and
    $\nabla_{\rm ad}\simeq 0.28310$ adopted by \protect\cite{2014MNRAS.445.3592P}
    to compare their predictions with those of a solar calibrated
    MLT.}
   \label{fig:sol}
\end{figure}

We show in Figs. \ref{fig:bubble_vb_rb} and \ref{fig:bubble_vb_rb_2} the
solutions of the bubble motion. When $\delta \rho < 0$ (continuous lines), the
bubble is rising and it reaches an asymptotic velocity, while $\omega =
\rho/\rho_b$, $r_b$, and $\log \frac{R}{R_0}$ increases continuously with
time. The value of the asymptotic velocity can be derived the following
way. When $\omega \gg 1$, eq. \ref{eq1} becomes
\begin{equation}
 \dot{v}_b =  \frac{2}{\gamma}  - \frac{1}{3\gamma} v_b^2.
\end{equation}
The asymptotic velocity corresponds to $\dot{v}_b = 0$, which leads
to $v_b^\infty = \sqrt{6}$. In physical units, this corresponds to $\sqrt{6}
c_s$. This value is shown as a horizontal dashed line in the left panels of
Fig. \ref{fig:bubble_vb_rb}. The asymptotic velocity is supersonic, which is
not consistent with the underlying assumptions of the theory. Therefore, it is
clear that this asymptotic velocity \emph{cannot} be used to compute a
convective flux.

The timescale on which the asymptotic velocity is reached depends only weakly
on the magnitude of the initial density perturbation, but it depends strongly
on the superadiabaticity. The smaller the superadiabaticity, the longer it
takes to reach the asymptotic velocity.  For the largest superadiabaticity
explored here, $\Delta \nabla = 10^{-1}$, the bubble expanded by a factor
$\sim 10$ and traveled a distance $\sim 10 H_p$ when it reaches the asymptotic
velocity. For $\Delta \nabla = 10^{-3}$, the bubble expanded by a factor $\sim
10 - 100$ and traveled over roughly $10^3 H_p$. For comparison, the number of
pressure scale height in the entire Sun is roughly 30.  As a conclusion, it is
clear that the time integration has to be stopped at some moment to make sure
that the velocity of the bubble remains subsonic and that the bubble did not
travel out of the convective region.

When $\delta \rho > 0$ (dashed lines), the bubble sinks in the
stratification. As a result, it contracts, and the magnitude of the velocity
increases with time. We find that two different outcomes are obtained: the
velocity diverges linearly in time for $\Delta \nabla = 0.1$, and the velocity
diverges at a finite time for $\Delta \nabla = 10^{-3}$.

When the superadiabaticity is large enough ($\Delta \nabla = 0.1$ in our case,
see Fig. \ref{fig:bubble_vb_rb}), $\omega = \rho / \rho_b$ decreases rapidly
as the bubble becomes more and more denser than its surrounding. When $\omega
\ll 1$, eq. \ref{eq1} becomes
\begin{equation}
\dot{v}_b = - 1 / \gamma.
\end{equation}
In physical units, this correspond to
\begin{equation}
  \label{eq:vb_downward1}
\dot{v}_b = - g.
\end{equation}
As nothing in the theory prevents the bubble to stop contracting,
its radius goes to zero and the bubble falls under the action of gravity alone
(free-fall). Its velocity diverges, and it becomes rapidly supersonic.

When the superadiabaticy is small enough ($\Delta \nabla = 10^{-3}$ in our
case, Fig. \ref{fig:bubble_vb_rb_2}), $\omega=\rho / \rho_b$ does not
decrease quickly enough, and the increase in the velocity magnitude now
results in the second term in eq. \ref{eq1} to be the dominant one. In this
case, Eq \ref{eq1} can be written as:
\begin{equation}
  \dot{v}_b = - C v_b^2,
\end{equation}
where $C$ is positive and can be considered constant in time. This gives immediately
\begin{equation}
  \label{eq:vb_downward2}
  v_b(t) = \frac{1}{- v_b^0 + C t},
\end{equation}
where $v_b^0$ is the (absolute) value of the bubble velocity at the moment
where the buoyancy force becomes negligible. One sees from
eq. \ref{eq:vb_downward2} that the bubble velocity diverges at $t = v_b^0 /
C$. This is a remarkable result that at a first sight may look surprising, yet
it can be understood in a very easy way and shows how unphysical the
predictions from the theory are. In the extreme case of a bubble moving
adiabatically in an adiabatic thermal stratification ($\Delta \nabla = 0$) the
buoyancy mass and the density contrast remain constant. In this case the
first term in the acceleration equation remains constant while the second one
increases as the bubble increases its speed. Once the second term becomes
dominant the bubble will contract extremely fast, shrinking to a point in a
finite timescale. Note that, as the density contrast remains constant to its
initial value $\omega=\omega(t=0)$ this means that at each time the bubble has
sunk deep enough so that its new density $\rho_b(t)$ follows that of the
background ($\rho(r_b(t))$). In particular this implies that when $R$ reaches
$R=0$ the bubble has already sunk to an infinite depth.

A particularly interesting conclusion that arises from the solution of the
motion of the adiabatic bubble is that there is no regime in which the
acceleration of the bubble fulfills the key eq. [41] of
\cite{2014MNRAS.445.3592P}. Not only ${v_b}^2\neq -\dot{v}_b R$ but, as shown
in the bottom right panels of Figs.  \ref{fig:bubble_vb_rb} and
\ref{fig:bubble_vb_rb_2}, the ratio ${v_b}^2/(\dot{v}_b R)$ changes over
orders of magnitude during the motion of the bubble. This is a very strong
result as this approximation is key in the derivation of the convective flux in
their work.

Finally, the previous results show that the theory cannot be
used to describe the motion of the bubble at all times. The time integration
has to be stopped when either one of the quantity $v_b$, $R$, $r_b$ reach a
value where the underlying assumptions of the theory cannot be verified
anymore.

\section{Discussion and concluding remarks}
\label{conclusion}

In the previous sections we have addressed the theory of convection
presented by \cite{2014MNRAS.445.3592P}. As discussed in section
\ref{mistakes} their theory is both a local and a time-independent
theory of convection, in the usual sense. In addition we have shown
that serious mathematical inconsistencies affect the derivation of the
final equations in \cite{2014MNRAS.445.3592P}, and that the key
physical assumption of a rapidly expanding bubble ($v_b/\dot{R}\ll 1$)
is in stark contradiction with the local and subsonic approach adopted
by the authors which requires $R/H_P\ll 1$. Yet, as we have shown in
sections \ref{eq_motion} and \ref{sol_motion}, it is possible to solve
the dynamics of the bubble consistently under the main physical
assumption of \cite{2014MNRAS.445.3592P}, i.e. assuming a
differentiated bubble moving in a potential flow. The detailed
analysis of the resulting solutions for the evolution of the bubble
show a very unphysical behavior. This is not a surprise, as potential
flows are known to be a far-fetched idealization of real fluids.
Indeed, it is known since d'Alembert that potential flows predict zero
drag, in strong contradiction with experience. This is the famous
``d'Alembert paradox'' \citep{1768dAlembert}. Potential flows are
popular in text books because they lead to analytically tractable
problems. However, potential flows are rarely achieved in real day
life, and they are mainly of academical interest
\citep{1964flp..book.....F}. In fact, the d'Alembert paradox shows
that the real flow around a body is not potential. Therefore, it is
clear that the assumption of a potential flow has the drawback that
the resulting theory will lack the drag that the fluid exerts on our
bubble.  Also, there is no physical reason to assume that a flow
  will remain potential even if that is the initial condition. 
The theory will necessarily be incomplete.  In addition, it is worth
noting that the relation between the acceleration, velocity and radius
of the bubble derived by \cite{2014MNRAS.445.3592P},
${v_b}^2=-\dot{v}_bR$, does not exist in the detailed solution of the
equations and is wrong by many orders of magnitude. This is important
because this relation is used to derive the expression for the
convective flux, which is key in their derivation of a convection
theory.

All the previous points indicate that no accurate description of stellar
convection can arise from the approach proposed by
\cite{2014MNRAS.445.3592P}. Yet, the authors claim that their theory is able
to reproduce the solar predictions of a sun-calibrated MLT. While this claim
looks surprising in view of the previous discussion, a closer inspection shows
that there is no such agreement. In fact, in their table 1 the authors quote
as a good agreement that their prediction for the temperature gradient
$\nabla$ differs in only $1.7\times 10^{-4}$ with the one predicted by the
sun-calibrated MLT. While this difference might look small at first sight, it
is a rather large discrepancy. The authors have chosen to compare their theory
in a regime that is still very close to adiabatic convection ---as can be seen
from the fact that the convective flux is 6 orders of magnitude larger than
the radiative one quoted in their table 1. The relevant prediction for a
convection theory is the degree of superadiabaticity
$\Delta\nabla=\nabla-\nabla_{\rm ad}$. As seen in Fig \ref{fig:sol} the
superadiabaticity in that region of the solar convective zone is between 
$\Delta\nabla\sim 6\times 10^{-5}$ and $3\times 10^{-4}$ ---either\footnote{It
  should be noted that the ``solar model'' adopted by
  \cite{2014MNRAS.445.3592P} is not a proper solar model, as the value of
  $\nabla$ at $R=0.98 R_\odot$ is not the correct one, see Fig
  \ref{fig:sol}. Also, Fig. 6 of \cite{2014MNRAS.445.3592P} shows the
  convective flux dominating down to $R=0.5R_\odot$, where the actual sun has
  no convective zone.} at $r=0.98
R_\odot$ or at the layer where $\nabla_{\rm ad}\sim 0.2831$. Therefore, the agreement for $\Delta\nabla$ between
the theory of \cite{2014MNRAS.445.3592P} and the sun-calibrated MLT is not
good, at best within an order of magnitude.

The study of the behavior of the dynamics of the expanding bubble forces us to
conclude that no improvement of stellar models can be expected from the
approach presented by \cite{2014MNRAS.445.3592P}. Indeed, the approach adopted
by the authors is too simplistic, beside the inaccuracies discussed in section
\ref{mistakes}, to be an accurate description of stellar convection.  Since 20
years, numerical simulations of stellar convection have shown that the flow
exhibits convective plumes, which are large scale, coherent structures that
emerge from the driving region and are able to propagate over significant
distance before loosing their identities \citep{stein_topology_1989,cattaneo_turbulent_1991,brummell_turbulent_1996,porter_three-dimensional_2000,brummell_penetration_2002-4,2013ApJ...769....1V}. The stratification has an important
role in stellar convection, as it breaks the symmetry between upflows and
downflows. For the case of stellar envelopes, where convection is driven by
cooling at the photosphere, convective plumes propagate downwards, and are
surrounded by a much broader and slower upflow. This is a result of mass
conservation. Convective plumes are seen to maintain their coherence over long
distances, i.e. larger than the pressure scale-height, and they are
responsible for the non-local character of convection. Furthermore, it is
known from numerical simulations that convective plumes contribute to energy
transport not only through the heat that they carry (enthalpy flux), but also
through their kinetic energy.  Due to the large Reynolds numbers that
characterize stellar hydrodynamics, convective plumes induce shear
instabilities as they propagate in the surrounding. As a result, they do not
have a very definite surface, nor a definite shape, as they continuously mix
with the surrounding. In some cases, this can reinforce the plume, as it
entrain more mass and becomes stronger. In some other cases, it can lead to a
destruction of the plume as it gets fully mixed with the surrounding, a
phenomenon called ``detrainment'' \citep{rieutord_turbulent_1995,rast_compressible_1998,clyne_interactive_2007}.
To be an improvement, future theories of stellar convection should take into account the non-local transport by convective plumes \citep{spruit_convection_1997, belkacem_closure_2006, 2015arXiv150403189B}.

The picture adopted in \cite{2014MNRAS.445.3592P} ignores
much of what has been learned from previous theoretical studies of stellar
convection. Although the ``bubbles'' which constitute the basis of
\cite{2014MNRAS.445.3592P} theory could be at first be identified as
representing convective plumes, it is clear the picture adopted by the authors
is too limited to really account for the observed properties of convective
plumes:
\begin{enumerate}
\item the authors adopt a local approach, in which the bubble size is
  restricted to length-scales smaller than the pressure scale-height, and in
  which both the dynamics of the bubble, as well as the predicted temperature
  gradients, only depend on the properties at each stellar layer;
\item the authors assume that motions are subsonic. This a valid approximation
  for the deep interior, where convection is efficient. However, close to the
  photosphere, the Mach number can be very large so that the flow cannot be
  considered as subsonic. There, one has no other choice than to consider the
  fully compressible equations of hydrodynamics.  For instance, at the photosphere the ram pressure of the fluid is large 
enough to modify hydrostatic equilibrium. This effect, which is 
described as due to a ``turbulent pressure", is neither taken into
account in MLT nor in the approach of  \cite{2014MNRAS.445.3592P};
\item the authors assume that the bubble has a well-delimited surface, along
  which the surrounding material is flowing. This picture is not able to
  account for shear instabilities that develop at the head of convective
  plumes, which of course have no definite surface. As mentioned previously,
  the way plumes entrain/detrain with the surrounding medium is key in setting
  their lifetime. This is the very reason why a mixing length is included
  ad-hoc in the MLT picture.
\end{enumerate}


Therefore, a non-local and time-dependent theory of stellar convection is still lacking,
hampering progress in stellar physics. Better predictions for the structure of superadiabatic layers
are required for asteroseismological studies. This requires to take into account compressibility (Mach numbers are of the order of one), and non-local effects due to plumes. State-of-the-art numerical simulations of photospheric convection are the most promising way to move beyond a MLT description of these layers. In the deep interior, although the thermal structure is know (the stratification is essentially adiabatic), a non-local theory of convection is needed to model the structure of the boundary layer between convective and stably stratified regions. This is timely, as both the extent and the efficiency of the mixing can now be probed with asteroseismology \citep{2011A&A...530A...3C,2015MNRAS.452..123C}.
A theory of convection remains elusive as it is an outstanding challenge to capture the richness of the phenomenon into a mathematical description. Furthermore, the current stellar structure equations offer a too restricted framework to do better than MLT-like, local descriptions of convection. Progress in this challenging field will likely result from physical insight gained from numerical simulations, a complete re-thinking of the stellar evolution equations (e.g., with stellar evolution codes evolving toward mean-field hydrodynamics), and the use of the increasing quantity and quality of observational data available to constrain theoretical models. Unfortunately, the work by \cite{2014MNRAS.445.3592P} does not provide any useful foundation for the success of this challenging, but necessary, enterprise.




\section*{Acknowledgements}
M3B is supported by a fellowship for postdoctoral researchers from the
Alexander von Humboldt Foundation. This research was supported by the Munich
Institute for Astro- and Particle Physics (MIAPP) of the DFG cluster of
excellence "Origin and Structure of the Universe". Part of this work was funded by
the European Research Council through grants ERC-AdG No. 341157-COCO2CASA.





\bibliographystyle{mnras}




\appendix

\section{Relationship between $P_b$ and $P_\infty$}
\label{app}
It is also interesting to analyze how the predictions of
eq. \ref{eq:presion} for the connection between the pressure inside
the bubble $P_b$ and the pressure of the fluid far away from the
bubble $P_\infty$. For this some physical comment about the hypothesis
of the spherical symmetry of the bubble is due. It is clear from
eq. \ref{eq:field} that the pressure on the surface of a sphere moving
within a fluid is \emph{not} constant. In the absence of any other
forces this differential forces will deform the shape of the
bubble. Then, the hypothesis of a spherically symmetric bubble at all
times is equivalent to assume that forces on the surface of the bubble
are able to balance the differential forces and keep a spherical shape
but do not prevent the sphere form expanding (i.e. a mechanical
constraint).

In order to obtain the link between the pressure inside the bubble $P_b$
(assumed to be filled with a homogeneous fluid) and the pressure in the fluid
we can analyze the energy transferred during an adiabatic \emph{spherically}
symmetric expansion $\mathrm{d}V_b$. Under the
assumption that the surface forces only act to prevent the departure from
spherical symmetry we can then write that the work done by the sphere has to
be equated by the energy received by the rest of the fluid, i.e.
\begin{equation}
P_b \mathrm{d}V_b =-\int_{\partial V} P \mathbf{n'}\cdot \mathrm{d}\mathbf{r'} dS
\end{equation}
as $\mathrm{d}\mathbf{r'}$ and $\mathbf{n'}$ are parallel during and spherical
expansion we can then write
\begin{equation}
P_b =-\frac{\int_{\partial V} P dS}{4\pi R^2}.
\end{equation}
Then, using eq. \ref{eq:presion} we find that
\begin{equation}
P_b = -\frac{\int_{\partial V} \left(\partial_t \psi + \frac{\mathbf v^2}{2} +  \Phi_g\right) dS}{4\pi R^2}.
\end{equation}
Which provides a link between the pressure inside the bubble and the state of
the surrounding material.  Now we can make of of the choice of $C'=0$ for the
relation the between pressure and the gravitational potential far away from
the sphere and write $\Phi_g=-g (z-r_b)-P_\infty^t/\rho$ ---where $P_\infty^t$
is the pressure far away from the bubble at the layer $z=r_b(t)$.
\begin{eqnarray}
 P_b =-\frac{1}{4\pi R^2}\left( \int_{\partial V} \partial_t \psi dS \right. &+&
  \int_{\partial V} \frac{\mathbf v^2}{2} dS +  \nonumber\\
+ \int_{\partial V}  (-g z') dS &+& \left.\int_{\partial V} P_\infty^t   dS \right)
\end{eqnarray}
Performing the integrations we find that
\begin{equation}
  P_b =P_\infty^t - \frac{\rho}{2} \dot{R}^2
- \frac{\rho}{2} {v_b}^2 \frac{1}{4\pi R^2}\int \frac{\sin^2\theta}{2} dS
- \ddot{R} R \rho - 2\rho \dot{R}^2
\end{equation}
Using the definition of the adiabatic sound speed we can now replace
$\rho=\Gamma_1 P_\infty/{c_s}^2$ to get
\begin{eqnarray}
  P_b =P_\infty^t &-&  \frac{\Gamma_1 P_\infty}{{c_s}^2} \dot{R}^2
- \frac{\Gamma_1 P_\infty}{{c_s}^2} {v_b}^2 \frac{1}{4\pi R^2}\int
\frac{\sin^2\theta}{2} dS \nonumber\\
&-& \frac{\Gamma_1 P_\infty}{{c_s}^2} \ddot{R} R  - \frac{\Gamma_1 P_\infty}{{c_s}^2} 2 \dot{R}^2.
\label{eq:LTA}
\end{eqnarray}
Eq. \ref{eq:LTA} shows the fact that  when $\dot{R}\ll
c_s$, $v_b\ll c_s$ and $\ddot{R}R \ll c_s^2$
the pressure inside the bubble can be considered to be equal to that of the
medium far away from the bubble $P_b \simeq P_\infty^t$, as it is well known
for subsonic flows.



\bsp	
\label{lastpage}
\end{document}